\documentclass{webofc}
\usepackage[varg]{txfonts}   
\newcommand{\cme}{\sqrt{s}}

\newcommand{\pdrap}{\lvert \eta \rvert}
\begin{document}
\title{Effects of phase space variables on oscillations of modified combinants}
%
%

\author{\firstname{H.W.} \lastname{Ang}\inst{}\thanks{\email{ang.h.w@u.nus.edu}}
        \and
        \firstname{Z.} \lastname{Ong}
             \and
        \firstname{P.} \lastname{Agarwal}
        \and
        \firstname{A.H.} \lastname{Chan} 
        \and
        \firstname{C.H.} \lastname{Oh} 
}

\institute{Department of Physics, National University of Singapore
          }

\abstract{%
  It has been shown recently that additional information can be obtained from charged particle multiplicity distribution by investigating their modified combinants $C_j$, which exhibit periodic oscillatory behaviour. The modified combinants obtained from experimental data can be expressed in a recurrent form involving the probability of obtaining $N$ charged particles $P(N)$, scaled by the void probability $P(0)$. The effects of various experimental observables such as $|\eta|$, $p_T$ and centre-of-mass collision energy $\sqrt{s}$ on the oscillations of $C_j$ will be discussed.
}
\maketitle
\section{Introduction}
\label{intro}
The "inner-space outer space connection" was one of the postulates conceived in the early 1990s to explain the numerous connections between the fields of particle physics and cosmology. One of these connections manifests in the form of multiplicity distribution (MD) of charged hadrons to the distribution of observed galaxies, both of which are products of the process of hadronization. Another similarity lies in the study of void probability $P(0)$ \cite{bib.voidprob1}, which represents the probability of observing no galaxies within a certain region of space. In the context of multiparticle production, $P(0)$ represents the probability of observing no hadrons within a certain region of phase space. 

In this work, we introduce the notion of modified combinant $C_j$ to be used in the analysis of MD. The $C_j$'s derived from experimental data have been observed to undergo periodic oscillations as a function of rank $j$. Their relationship to $P(N)$, as well as the effects of pseudorapidity $\eta$, transverse momentum $p_T$ and centre-of-mass collision energy $\cme$ on their oscillation will be explored.

\section{Relationship between $C_j$ and $P(N)$}
\label{sec-2}
Apart from the generating function and probability function, information in a statistical distribution can also be expressed in a recurrent form involving only adjacent values of probability $P(N)$

\begin{equation}
    (N+1)P(N+1) = g(N)P(N).
\label{eqn.recurrent.P(N)}
\vspace*{0.5cm}
\end{equation}A linear form of $g(N)$, though useful, is somewhat limited in the sense that any $P(N)$ is only related to the value of $P(N-1)$. This constraint seems unnecessarily restrictive. In \cite{bib.Cj.origin}, it is proposed that Eqn (\ref{eqn.recurrent.P(N)}) be modified in order to factor effects of all $P(j)$'s for $j < N$

\begin{equation}
    \langle N \rangle C_j = (j+1)\left[ \frac{P(j+1)}{P(0)} \right] - \langle N\rangle \sum^{j-1}_{i=0}C_i \left[ \frac{P(j-i)}{P(0)} \right]
    \label{eqn.Cj}
    \vspace*{0.5cm}
\end{equation}

The coefficients $C_j$ in Eqn (\ref{eqn.Cj}) are the modified combinants. In this form, it is clear that $C_j$ acts as the weight that determines the relative contributions of all smaller $P(N-j)$'s to the value of $P(N)$. In some sense, one can interpret $C_j$ as the "memory" that the $P(N)$ term has of all lower multiplicity terms.

\section{Dependence of $C_j$ oscillations on phase space variables}
\label{sect.phase.space.dep}
In high-energy collision experiments, the phase space is typically characterised by three variables - $\eta$, $p_T$ and $\cme$. Each of these variables contain information of the characteristics of the collision, which affect the probability of producing a certain number of particles. On this note, we will study the effects of each of these three variables on the oscillation of $C_j$.

\subsection{Dependence on $\eta$}
\label{subsect.eta.dep}

\begin{figure}[t]
\centering
     \begin{minipage}{0.45\linewidth}    	
    	\includegraphics[width=1.0\linewidth]{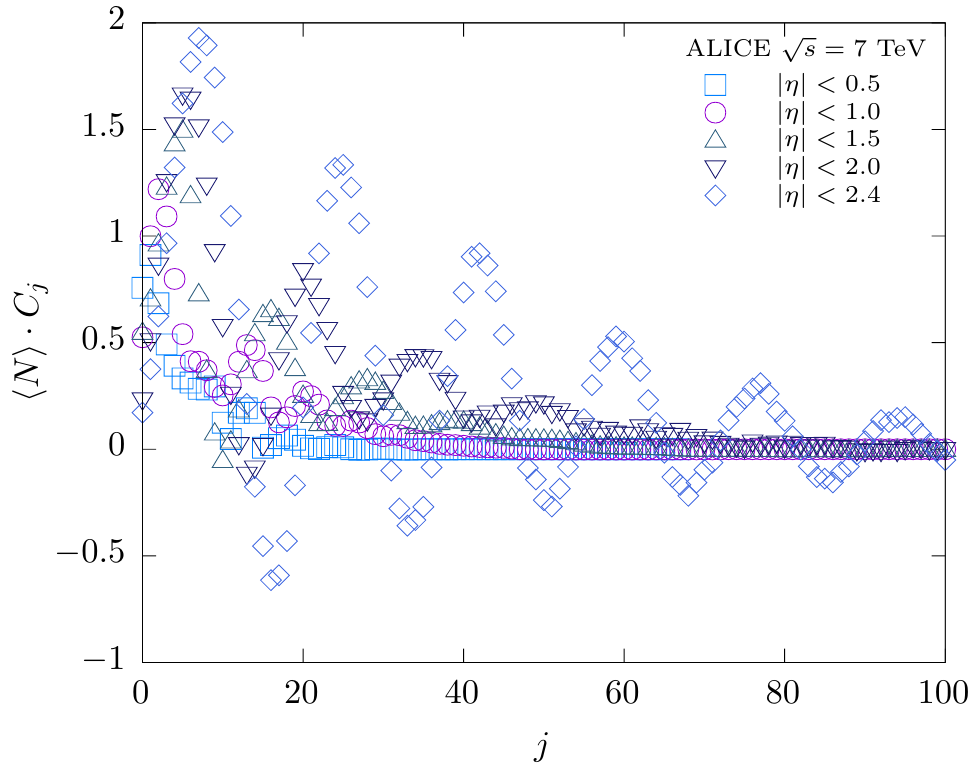}
    \end{minipage} \hfill
	\begin{minipage}{0.45\linewidth}		
        \includegraphics[width=1.0\linewidth]{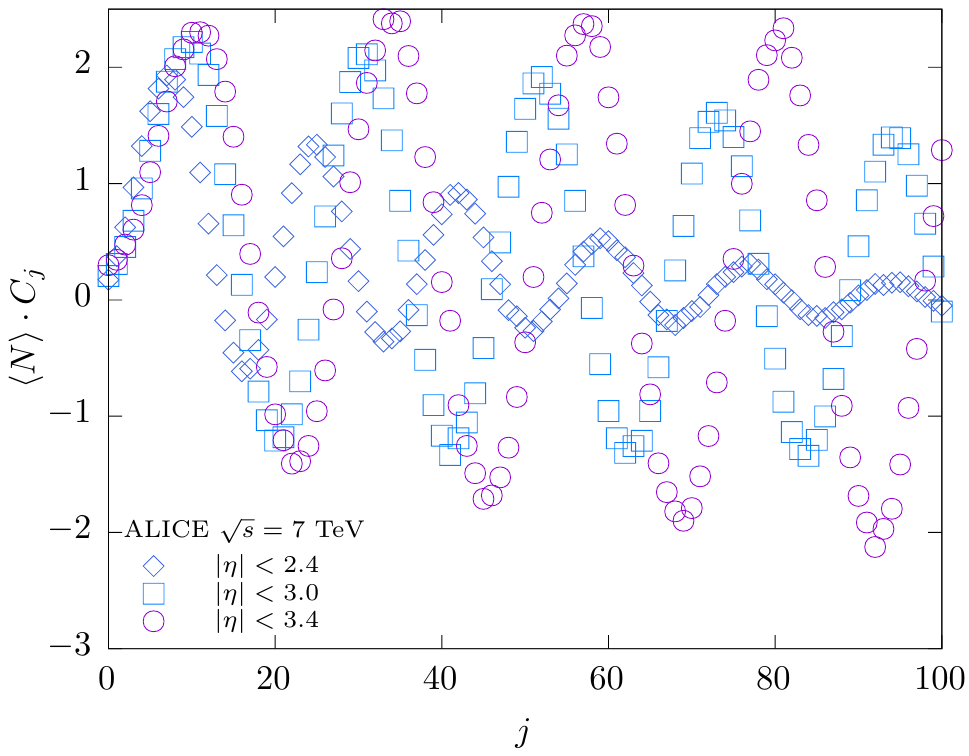}
	\end{minipage}
\caption{Plot of $C_j$ derived from ALICE data. {\bf Left panel:} The plot is made with MD data up to $\pdrap < 2.4$. The amplitude increases with wider pseudorapidity interval which decays with increasing $j$. {\bf Right panel:}  The plot made with data up to $\pdrap < 3.4$ shows that the amplitude decay stagnates at $\pdrap < 3.0$ and increases with increasing $j$ at $\pdrap < 3.4$.}
\label{fig.eta}       
\end{figure}

The pseudorapidity variable, denoted by $\eta$, characterises the angle relative to the beam axis. The prefix "pseudo" stems from the fact that it is used as a proxy for rapidity (typically denoted by $y$). The experimental calculation of rapidity involved knowing the longitudinal component of a particle's momentum $p_z$ and its energy $E$, quantities which are not easily available. As such, the pseudorapidity variable $\eta$ defined in place as a function of $\theta$, the angle between the forward beam direction and the particle's 3-momentum. This is a more experimentally accessible quantity and in high-energy physics, where the momentum is much larger than the rest mass of a particle, $\eta \approx y$.

The behaviour of $C_j$ plotted as a function of $j$ from experimental data published by the ALICE collaboration \cite{bib.ALICE8TeVwide,bib.ALICE8TeV} is shown in Fig \ref{fig.eta}. 
It is evident that both the period and amplitude of the oscillations increase with the pseudorapidity interval. Comparing the left and right panels of Fig. \ref{fig.eta}, one observes that the rate of amplitude decay decreases at larger pseudorapidity interval. In fact, the amplitude seems to remain constant at $\pdrap < 3.0$ before reversing and increases for $\pdrap < 3.4$.

\subsection{Dependence on $p_T$}
\label{subsect.pt.dep}

\begin{figure}[t]
\centering
     \begin{minipage}{0.45\linewidth}    	
    	\includegraphics[width=1.0\linewidth]{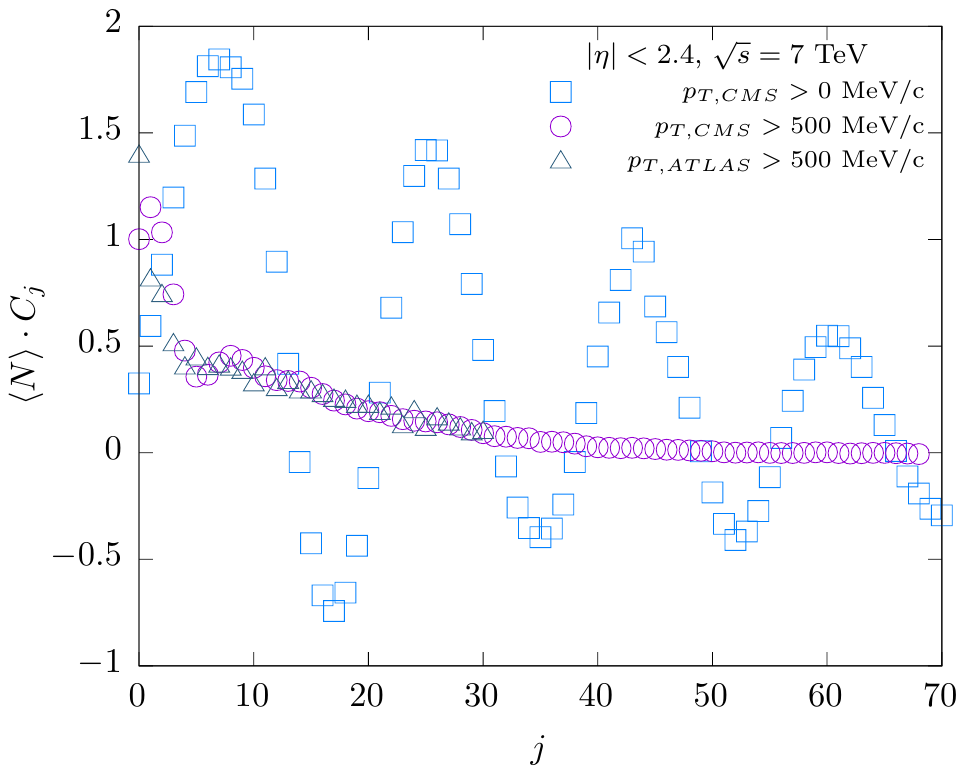}
    \end{minipage} \hfill
	\begin{minipage}{0.45\linewidth}		
        \includegraphics[width=1.0\linewidth]{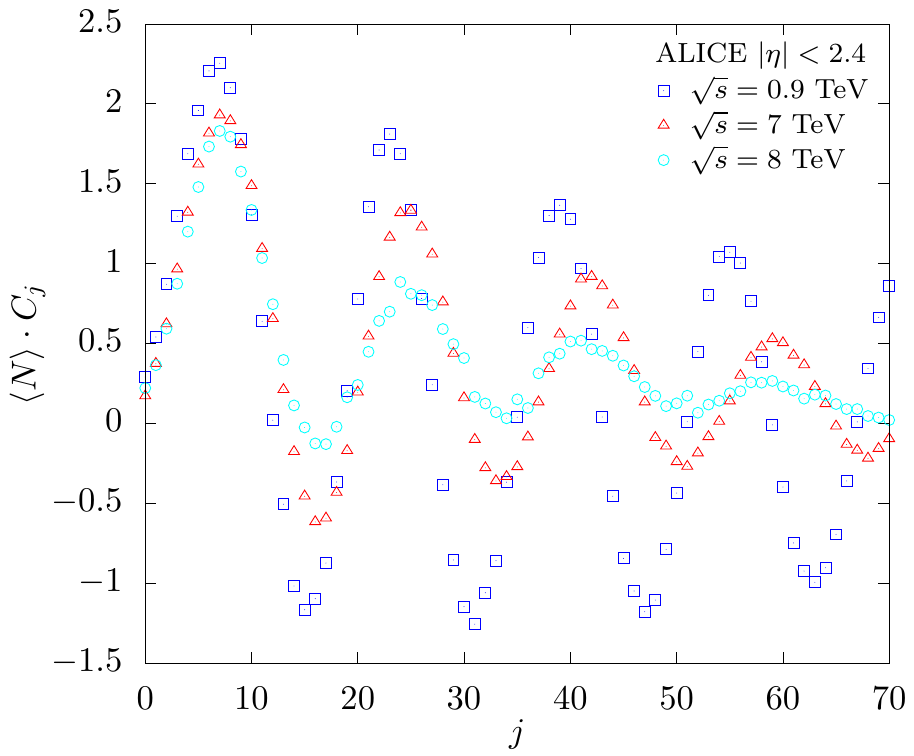}
	\end{minipage}
\caption{\textbf{Left panel:} The $C_j$ oscillations across different transverse momentum cuts from data in \cite{bib.CMS.run1}. Once again, data from a larger phase space volume of $p_T > 0$ MeV/c exhibits more significant oscillation. \textbf{Right panel:} $C_j$ plotted across various centre-of-mass energies at the same pseudorapidity interval of $\pdrap < 2.4$ from ALICE data.}
\label{fig.pt}       
\end{figure}

In high-energy collisions, transverse momentum, or $p_T$, is the component of the particle's 3-momentum that is perpendicular to the beam axis. It provides an indication of the "hardness" of the underlying interaction in the collisions. 

In general, harder collisions give rise to particles with higher $p_T$ while particles resulting from softer collisions possess lower $p_T$. Hard (or short-distance) interactions can be accurately described by perturbative QCD (pQCD) in increasing powers of the coupling constant $\alpha_s < 1$ in this regime. Unfortunately, such perturbative techniques fail in describing the physics of soft (or large distance) interactions, which corresponds to $\alpha_s > 1$.

A reference to a plot of $p_T$ vs $N$ (such as that in \cite{bib.CMS.run1}) will show that soft interactions produce majority of the detected secondary particles as compared to the hard interactions. This obliges the introduction of various phenomenological models to augment pQCD in providing a complete picture of the underlying physics.

Such a characteristic makes the study of the dependence of $C_j$ oscillations on $p_T$ of interest. In left panel of Fig. \ref{fig.pt}, the pseudorapidity interval\footnote{The data from the ATLAS collaboration is taken over a slightly wider interval of $\pdrap < 2.5$ compared to $\pdrap < 2.4$ from CMS. However, as observed in the left panel of Fig. \ref{fig.pt}, the slightly wider interval makes no practical difference at $p_T > 500$ MeV/c.} and centre-of-mass energy variables are kept constant to isolate the effects of changing $p_T$ on $C_j$. Oscillations are only observed for the experimental data extrapolated to $p_T > 0$ MeV/c from the CMS collaboration, which includes more secondary particles from the soft collisions.

\subsection{Dependence on $\cme$}
\label{subsect.cme.dep}
Centre-of-mass energy, denoted by $\cme$, provides an indication of the maximum amount of energy available for exchange in interactions during a high-energy collision. In a nutshell, collisions at a higher $\cme$ has a higher probability of producing more particles due to the mass-energy relation $E^2 = p^2 + m^2$. In this way, the multiplicity of the collisions is also a function of $\cme$.

Once again, in the right panel of Fig. \ref{fig.pt}, the pseudorapidity interval has been kept constant at $\pdrap < 2.4$ to isolate the effects of a changing $\cme$. It is observed that the lower the $\cme$, the larger the amplitude of $C_j$ oscillations, and the slower the rate of amplitude decay. However, there is no clear trend in the period of the oscillation with increasing energy.

\section{Discussion of results}
\label{sect.results}
It is clear that each of the three experimental variables discussed has effects on the oscillations of $C_j$, and by extension the fluctuation of "memory" on lower multiplicities. In Sect. \ref{subsect.eta.dep}, we see that the oscillations change violently with $\eta$. In general, the larger the $\pdrap$ interval, the larger the oscillations and the slower the oscillations decay. This seem to indicate that a larger region of available phase space corresponds to more violent oscillations in $C_j$. In a way, one can view information captured from a larger $\pdrap$ interval as being more indicative of the total available information of the underlying physics. This can simply be explained from the very fact that a wider $\pdrap$ interval captures more of the produced particles, as well as decays from leading particle.

The tentative hypothesis on larger available phase space leading to more violent oscillations is also supported by the plot in the left panel of Fig. \ref{fig.pt}. In this plot, the cut made at $p_T > 0$ MeV/c will undoubtedly capture more of the underlying produced particles than the cut made at $p_T > 500$ MeV/c, with a greater available phase space volume in the former. Once again, this leads to a more violent oscillation in $C_j$ with the increased phase space volume.

Interestingly, the dependence of oscillation on $\cme$, as depicted on the right panel of Fig. \ref{fig.pt}, does not corroborate with the hypothesis proposed above. If one were to restrict the definition of phase space variables to that involving the spatial (proxied by $\pdrap$) and momentum ($p_T$) variables, then the $\cme$-dependence should not be interpreted in the light of the hypothesis, which describes only dependence on phase-space variables.

This study elucidates the dependence of $C_j$ oscillations on availability of phase space, or extent of the available phase space volume. In general, the larger the available phase space, the more violent the $C_j$ oscillations. A tentative phenomenological model has been proposed to explain the existence of such oscillations in \cite{bib.multipart.prod}. Model description of the fluctuating $C_j$ is possible if one includes the binomial distribution as one of the two distributions to be compounded in a compound distribution. 

Comparison of the differences in the period of oscillation and increase in amplitude also leads to a proposal of using $C_j$ oscillations to distinguish between matter-matter and matter-antimatter collisions as suggested in \cite{bib.multipart.effects}.

\section{Conclusion}
The dependence of $C_j$ oscillations on $\eta$, $p_T$ and $\cme$ are partially explored in this work. It is seen that $C_j$ provides an alternate means to interpret the multiplicity data obtained from high-energy collision experiments under various conditions. As mentioned in section \ref{intro}, if the inner-space-outer-space connection holds true, $C_j$ may also be considered a potential candidate as a tool in the study of galaxy distribution in our universe. In the vast expanse of space, quantities such as the void probability $P(0)$ should be obtainable via observation and with it, calculations on $C_j$ based on observational data can be performed. 

The utility of $C_j$ as introduced in this work is far from complete. In \cite{bib.multipart.effects}, links between $C_j$ and various moments, as well as to the non-extensive parameter $q$ in Tsallis statistics \cite{bib.Tsallis}, have been made. Hence, there is much to be explored in the full range of utility of $C_j$, this work of which is only the tip of the iceberg.

\section*{Acknowledgement}
We (H.W. Ang, Z. Ong and P. Agarwal) will like to thank the NUS Research Scholarship for partially funding this study. One of us (H.W. Ang) will like to thank Prof. G.~Wilk, Prof Z.~Włodarczyk and Prof M.~Rybczyński for their critical discussions and comments.
%
%
%

\end{document}